\documentclass{PoS}

\usepackage{graphicx}
\usepackage{dsfont}
\usepackage{booktabs}
\usepackage{amsmath}

\title{Simulating the Random Surface representation of Abelian Gauge Theories}

\ShortTitle{Simulating the Random Surface representation of Abelian Gauge Theories}

\author{\speaker{Tomasz Korzec}\\
        Humboldt Universit\"at zu Berlin\\
        E-mail: \email{korzec@physik.hu-berlin.de}}

\author{Ulli Wolff\\
        Humboldt Universit\"at zu Berlin\\
        E-mail: \email{uwolff@physik.hu-berlin.de}}

\abstract{We present a Monte-Carlo algorithm for the simulation of 
the all-order strong coupling expansion of the $Z_2$ gauge theory. This 
random surface ensemble is equivalent to the standard
formulation, but allows to measure some quantities, like Polyakov loop
correlators or excess free energies, with an accuracy that could not have
been easily achieved with traditional simulation methods. One interesting 
application of the algorithm is the investigation of effective string 
theories.

\begin{flushright} HU-EP-13/70 \end{flushright}
\begin{flushright} SFB/CPP-13-99 \end{flushright}
}

\FullConference{31st International Symposium on Lattice Field Theory LATTICE 2013\\
                 July 29 - August 3, 2013\\
                 Mainz, Germany}

\begin{document}

\section{Introduction}
Monte Carlo algorithms that are based on all-order strong-coupling (or hopping-parameter) expansions
of statistical systems have been enjoying an increasing popularity ever since the invention of
the worm algorithm~\cite{Prokof'ev:2001zz}. For many systems, including nonlinear $O(N)$ $\sigma$
models~\cite{Wolff:2009kp,Janke:2009rb}, CPN models~\cite{Wolff:2010qz} and even some fermionic 
systems~\cite{Wolff:2008xa,Wenger:2008tq,Baumgartner:2011jw}, 
worm algorithms have been introduced
that show nearly no signs of critical slowing down and often allow to access interesting observables that 
are difficult to measure in the traditional formulation. In other cases models that suffer from a 
sign problem have been successfully treated by worm algorithms~\cite{Panero:2005iu,Endres:2006xu,Mercado:2013ola,Chandrasekharan:2012va}. 
An introduction to the topic and
further references can be found in recent reviews~\cite{Wolff:2010zu,Gattringer2013}.

For gauge theories the situation remains less clear so far. In the Abelian case, several algorithms based
on the all-order strong coupling expansion have been investigated~\cite{Sterling:1983fs,Azcoiti:2009md,Korzec:2010sh}.
It seems that a fast dynamics
near critical points is difficult to achieve, while other advantages of worm algorithms have not been
studied in detail. In this work we explore several worm-related methods for the simulation of the
$Z_2$ Gauge model and focus on the formulation of highly improved estimators for quantities that are
usually difficult to calculate accurately.

\section{Random Surface Representation of $Z_2$ Gauge Theory}
The simplest and oldest lattice gauge theory is Wegner's gauge invariant Ising model~\cite{Wegner:1984qt}
with plaquette action
\begin{equation}
   S[\sigma] = -\beta \sum_{x,\mu<\nu} \sigma(x,\mu)\sigma(x+\hat\mu,\nu)\sigma(x+\hat\nu,\mu)\sigma(x,\nu) \, .
\end{equation}
The sites of a hypercubic lattice are labeled by $x$, the links between sites $x$ and $x+\hat \mu$ by $(x,\mu)$
and $\sigma\in\{-1,+1\}$ is the gauge field.
We write the most general observable in terms of a ``defect network'' $j(x,\mu)\in\{0,1\}$
\begin{equation}
   O[j] = \prod\limits_{x,\mu} \sigma(x,\mu)^{j(x,\mu)} \, .
\end{equation}
Its ensemble average is given by $\langle O[j] \rangle = \frac{Z[j]}{Z[0]}$ with 
the generalized partition function
\begin{equation}
   Z[j] = \sum_{\{\sigma\}} e^{-S[\sigma]} \ O[j] \, .
\end{equation}
Expanding 
\begin{equation}
   e^{\beta\sigma_p} = \cosh(\beta) \sum_{n=0}^1 t^n \sigma_p^n, \qquad\qquad  t = \tanh(\beta)\, ,
\end{equation}
on each of the $N_p$ plaquettes of the lattice, allows to carry out the summations 
over the gauge fields and leads to the random surface representation
\begin{equation}
    Z[j] \propto \sum_{\{n\}} \delta(\partial n+j)\ t^{\sum\limits_{x,\mu<\nu} n_{\mu\nu}(x)} \, .
\end{equation}
The summation is over $n_{\mu\nu}(x) \in \{0,1\}$ on each plaquette and the configurations can be interpreted
as surfaces composed from $n=1$ plaquettes. Their (positive) weight depends on the total surface, given that
the constraint
\begin{equation}
   \delta(\partial n+j) \equiv \prod_{x,\mu} \delta_{j(x,\mu) + \sum\limits_{\nu>\mu} n_{\mu\nu}(x)+ n_{\mu\nu}(x-\hat\nu) \mod\ 2,\ 0}
\end{equation}
is satisfied. An intuitive geometric meaning is that only surfaces whose boundaries are given by $j$ contribute to the ensemble. In particular
only closed surfaces  enter into $Z[0]$.

We are mainly interested  in Polyakov line correlators
\begin{equation}
    O[j^{(\mathbf u,\mathbf v)}] = \pi(\mathbf u)\pi(\mathbf v) \, , \qquad \pi(\mathbf u)=\prod\limits_{u_0} \sigma(u, 0)\, .
\end{equation}
The first equation implicitly defines the defect network associated with two Polyakov lines at spatial 
positions $\mathbf u$ and $\mathbf v$.
Inspired by the successful worm algorithms for spin models, we consider the enlarged system
\begin{equation}
   \mathcal{Z} = \sum_{\mathbf u,\mathbf v} \rho^{-1}(\mathbf u - \mathbf v) \
                 Z[j^{(\mathbf u,\mathbf v)}]
                \ = \ \sum_{\{n\},\mathbf u, \mathbf v} \rho^{-1}(\mathbf u - \mathbf v) \
                 t^{\sum\limits_{x,\mu<\nu} n_{\mu\nu}(x)} \ \delta(\partial n+j^{(\mathbf u,\mathbf v)})\, .\label{largeens}
\end{equation}
The weight function $\rho\geq 0$, $\rho(0)=1$ is for the moment arbitrary and will be specified later.
Ensemble averages with respect to $\mathcal{Z}$ are denoted by double brackets $\langle\langle O \rangle\rangle$.

\section{Observables}
Observables in the original ensemble can be related to those in the enlarged one.
By design, the Polyakov line correlation function is one of the simplest cases
\begin{equation}
G(\mathbf x) = \langle \pi(\mathbf x) \pi(\mathbf 0) \rangle = 
   \rho(\mathbf x) \frac{\langle\langle\delta_{\mathbf x,\mathbf u - \mathbf v} \rangle\rangle}
                                        {\langle\langle\delta_{\mathbf u, \mathbf v} \rangle\rangle} \, .
\end{equation}
In contrast to the original formulation this estimator does not fluctuate in sign.
Moreover, if the chosen $\rho$ is an approximation of the expected result, the Monte-Carlo simulation calculates 
only the correction to this approximation, which in the best case depends only weakly on $\mathbf x$.

For many other observables only closed-surface configurations are relevant and it is convenient
to introduce an abbreviation for averages of these ``vacuum observables'' 
\begin{equation}\label{vacuumObs}
\langle\langle O[n] \rangle\rangle_0 \equiv \frac{\langle\langle O[n]\ \delta_{\mathbf u, \mathbf v}\rangle\rangle}
                                                 {\langle\langle \delta_{\mathbf u, \mathbf v}\rangle\rangle}
\end{equation}
The average plaquette is of this type
\begin{equation}
   \langle \sigma_p \rangle = \frac{\partial \log Z}{\partial \beta} = 
                              t+\frac{(t^{-1}-t)}{N_p}\Bigl\langle\Bigl\langle \sum\limits_{x,\mu<\nu}n_{\mu\nu}(x)\Bigr\rangle\Bigr\rangle_0\, .
\end{equation}
It is trivially related to the total surface of boundary free graphs and takes the role of 
internal energy.

In~\cite{Korzec:2012fa} it has been shown that twisted boundary conditions can be 
easily incorporated into the random surface representation.
The partition function of a system with twists $\gamma_{\mu\nu} \in \{-1,+1\}$ in the $\mu/\nu$ planes is denoted
by $Z_\gamma$. The periodic case, where all twists are one, is simply $Z$. Partition function ratios can
be calculated by
\begin{equation}
       \frac{Z_\gamma}{Z} = \Bigl\langle\Bigl\langle \prod\limits_{\mu<\nu} (\gamma_{\mu\nu})^{w_{\mu\nu}[n]}\Bigr\rangle\Bigr\rangle_0 \, ,
       \qquad
       w_{\mu\nu}[n] = \sum_{x|_{x_\mu=x_\nu=0}} n_{\mu\nu}(x)\quad  (\text{mod}\ 2) \, .
\end{equation}
A configuration has a wrapping number $w_{\mu\nu}$ for every $\mu/\nu$ plane.
Such ratios for all possible twists can be obtained from a single simulation. They are 
renormalized observables which due to their strong dependence on $\beta$ can be useful
to determine the critical coupling.

\section{Monte Carlo Updates}
The system with partition function $\mathcal{Z}$ is well suited for treatment with Monte-Carlo methods.
A configuration that satisfies the constraint has the positive weight
\begin{equation}
   W(n;\mathbf u, \mathbf v) = \frac{1}{\mathcal{Z}} \rho^{-1}(\mathbf u - \mathbf v) \
                   t^{\sum\limits_{x,\mu<\nu} n_{\mu\nu}(x)}
\end{equation}
In algorithms of Metropolis type a proposed change $(n;\mathbf u, \mathbf v) \to (n';\mathbf u', \mathbf v')$
gets accepted (if the proposal is symmetric) with probability
$\min\left[1,\frac{W(n';\mathbf u', \mathbf v')}{W(n;\mathbf u, \mathbf v)} \right]$.

\subsection{Cube Updates}
   The smallest change in a configuration that maintains the constraints has been introduced in a
   similar context in~\cite{Sterling:1983fs}. The proposal is to flip $n$ on the six faces of a 3D cube.
   First such a cube, i.e. $x$ and $\mu<\nu<\lambda$ is chosen (randomly or in an ordered fashion).
   The six plaquettes associated with it are
   ${\mathcal P} = \{(x,\mu,\nu),(x+\hat\lambda,\mu,\nu),(x,\mu,\lambda),(x+\hat\nu,\mu,\lambda),(x,\nu,\lambda),(x+\hat\mu,\nu,\lambda)\}$.
   The proposal $n(p)\to n'(p) = 1-n(p)$ for all $p\in{\mathcal P}$ is accepted 
   with probability $\min[1, \prod\limits_{p\in {\mathcal P}} t^{1-2n(p)}]$.
   Such an update does neither change the boundary of a surface nor its wrapping number, 
   and is therefore on its own insufficient to simulate the ensemble~(\ref{largeens}).
   In $D=3$ the cube update can be related to a local spin flip in the dual Ising
   spin system.  
   %For such updates critical slowing down with a dynamical exponent $z\sim 2$
   %was established~\cite{}.
   
\subsection{Cube-Cluster Update}
It is possible to formulate a single cluster algorithm based on the cube flips.
In a first step new bonds living on the plaquettes $b(p)\in\{0,1\}$ are chosen randomly. 
The value $1$ has probability $P(b(p)=1) = 1-t^{1-n(p)}$, in particular 
bonds through surface plaquettes are excluded. Next, a $3D$ cube $c_0 = (x,\mu,\nu,\lambda)$ is chosen randomly.
The cluster $C$ is defined as the maximal set of cubes such that for each cube $c\in C$
there exists a path from $c$ to $c_0$ that crosses only plaquettes with $b(p)=1$.
The entire path has to lie in the $3D$ subspace spanned by $\hat\mu, \hat\nu, \hat\lambda$.
In practice the bonds are determined on the fly, while the cluster is being constructed
(single cluster algorithm).

In a final step for every $c\in C$, the flip $n(p)\to 1-n(p)$ is carried out on all six plaquettes
associated with c. This amounts to flipping $n$ on the cluster boundary.
   
It is well known~\cite{Wolff:1988uh} that this algorithm nearly eliminates critical slowing down 
in $D=3$ with fluctuating twisted b.c.\footnote{The Ising spin system with periodic boundaries is mapped by a
duality transformation onto an Ising gauge model with fluctuating twisted b.c.}
The update remains valid for $D>3$, its efficiency in these cases has yet to be investigated.

\subsection{Polyakov Shift Updates}\label{psu}
In addition to the boundary preserving updates we introduce a worm update for the positions
of the Polyakov lines. A proposal is made to move the line at position $\mathbf u$ to 
to one of its $2(D-1)$ spatial neighbors $\mathbf u \to \mathbf u' = \mathbf u\pm\hat{\imath}$.
To maintain the constraints, the shift is accompanied by a flip of the whole ladder of temporal plaquettes 
with corners at spatial coordinates $\mathbf u$ and $\mathbf u\pm \hat{\imath}$. 
The acceptance probability is given
by $ \tilde p_{i,\pm} = \min \left[1, t^{L_0-2k(\mathbf u,\pm i)}\ \frac{\rho(\mathbf u-\mathbf v)}{\rho(\mathbf u'-\mathbf v)} \right]$,
where 
\begin{equation}   
   k(\mathbf x, i) = \sum\limits_{x_0=0}^{L_0-1} n_{0i}((x_0,\mathbf x)),\qquad 
                    k(\mathbf x, -i)\equiv k(\mathbf x-\hat{\imath},i)\, ,
\end{equation}
is the surface of the ladder before the flip.

This worm update is preceded by a random relocation of $\mathbf u$ and $\mathbf v$ in cases
when they are equal.

\subsection{Non-Rejecting Polyakov Shift Updates}
A Potential problem of the previous update is a decreasing acceptance rate when $L_0$ becomes large.
In cases where one is interested only in vacuum observables~(\ref{vacuumObs}) an idea from simulations of 
densely packed systems~\cite{Liu:2010uw} can be adopted. 
If $\mathbf u=\mathbf v$, the line at $\mathbf{u}$ is updated according to the previous worm algorithm.
If $\mathbf u\neq\mathbf v$ however, the update is changed. A move is always made (no accept/reject step), and the 
probability to move $\mathbf u$ to position $\mathbf u\pm\hat{\imath}\ $ is constructed from the acceptance probabilities
of section~\ref{psu} and is given by
\begin{equation}
   p_{i,\pm} = \frac{\tilde p_{i,\pm}}{2(D-1)A}, \qquad  A = \frac{1}{2(D-1)}\sum\limits_i(\tilde p_{i,+}+\tilde p_{i,-})\, .
\end{equation}
The modified Markov matrix is not in balance with the original probability 
distribution any more. Instead the partition function
\begin{equation}
            {\mathcal Z}_{\rm nr} = \sum_{\{n\},\mathbf u,\mathbf v} t^{\sum_p n(p)}\ \rho^{-1}(\mathbf u-\mathbf v)
               \ \delta(\partial n + j^{(\mathbf u,\mathbf v)}) \
               \left[\delta_{\mathbf u,\mathbf v} + (1-\delta_{\mathbf u,\mathbf v})A \right]
\end{equation}
is sampled. Fortunately the closed-surface configurations have the same relative weights as before.

\section{Results}
\subsection{At the Critical Point}
\begin{figure}[h]
\includegraphics[width=0.49\linewidth]{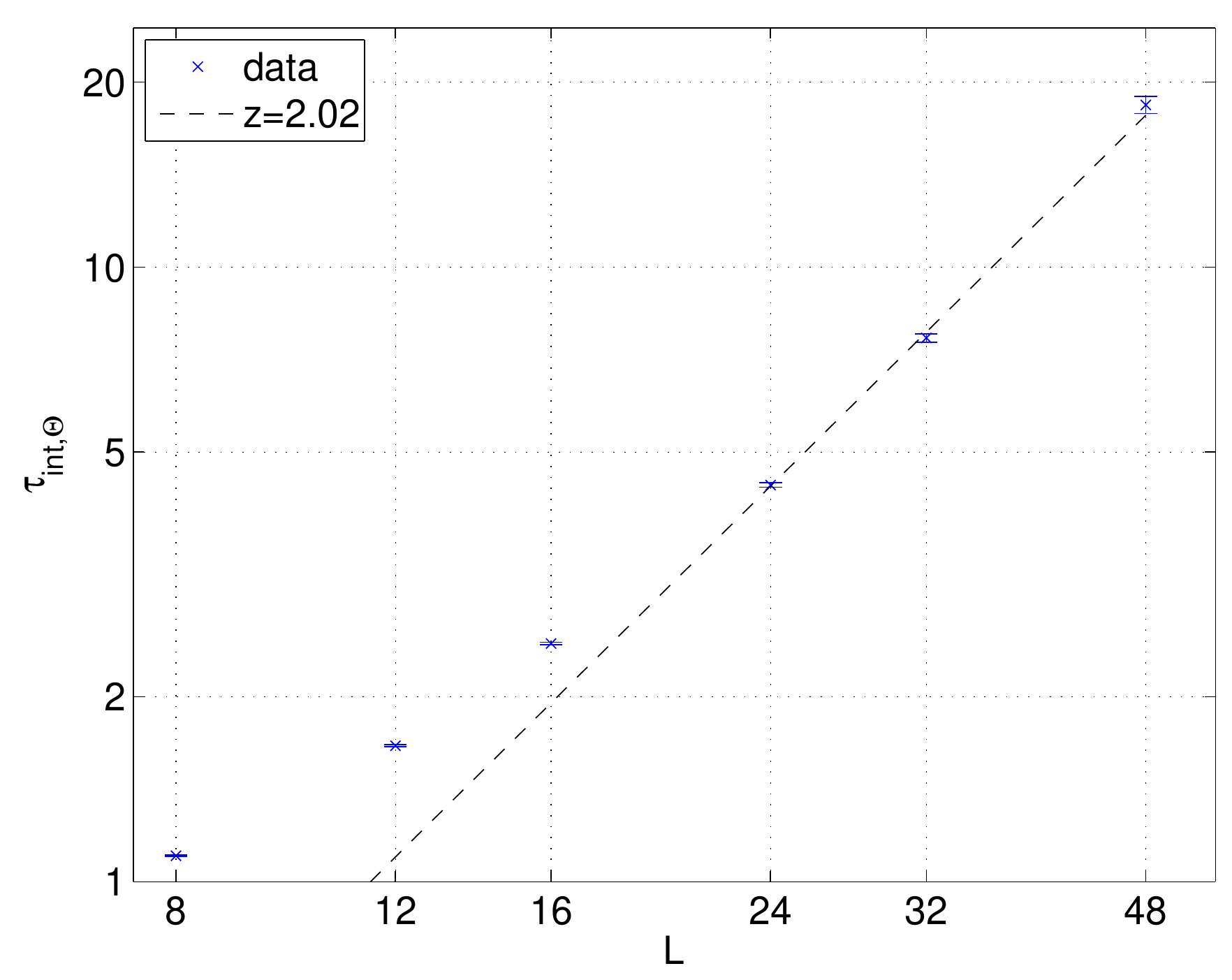}
\includegraphics[width=0.49\linewidth]{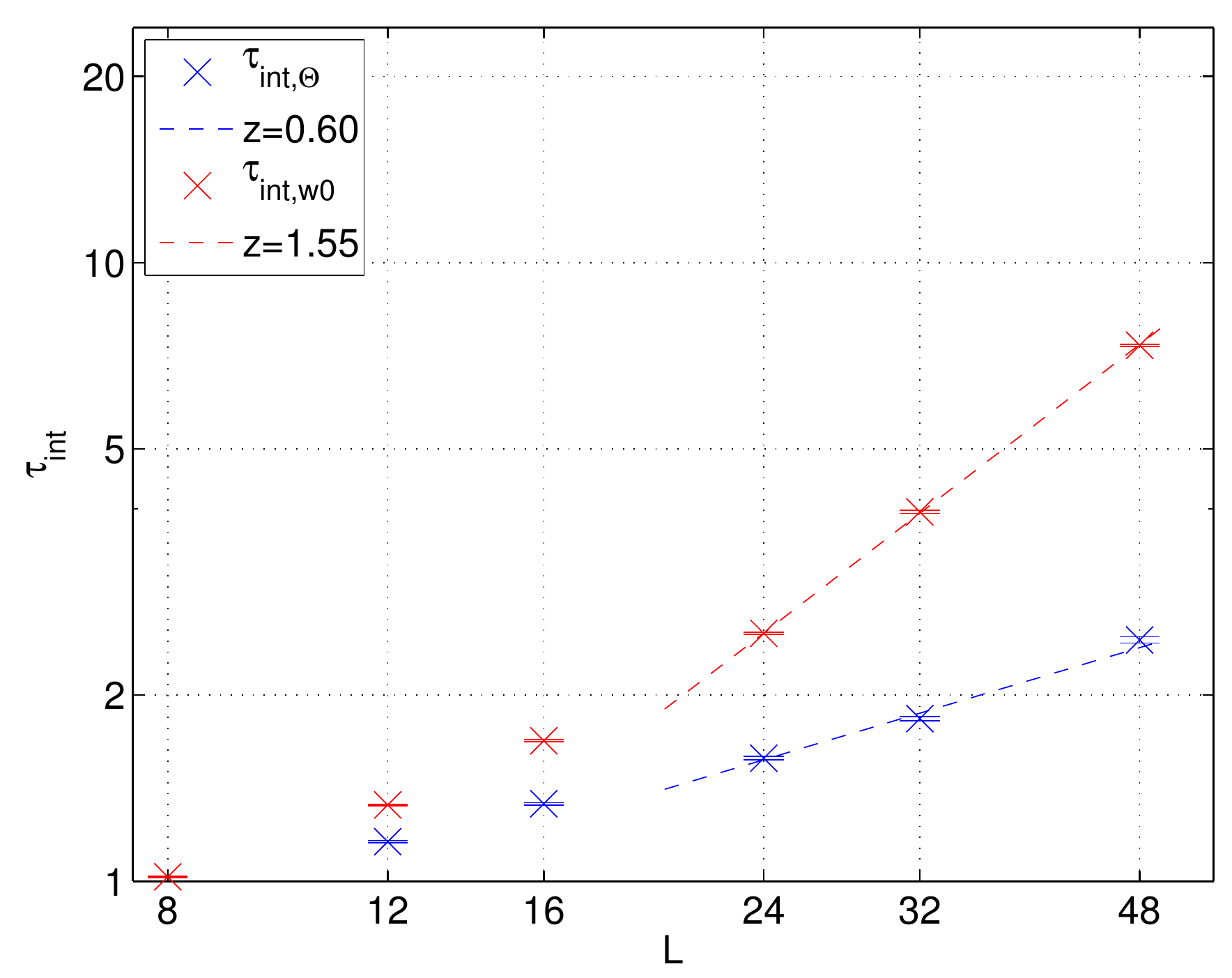}
\caption{Critical slowing down of the algorithm with and without cluster updates. $\Theta$ is the total
surface of vacuum graphs, while $w_0$ is the fraction of vacuum graphs with zero wrapping number in
all planes.}\label{FIGtauint}
\end{figure}

In a first set of simulations in D=3 dimensions, we set $\beta$ to its critical value\cite{Deng:2003wv,Hasenbusch2012},
$\beta_c \approx 0.76141346$. With a geometry where $L_0=L_1=L_2$ the lattice size
is varied, $L/a = 12,16,24,32,48$. The integrated autocorrelation times of various 
observables are measured. Figure~\ref{FIGtauint} shows the results in a double-logarithmic plot 
for two variants of the algorithm. A version in which cube updates and non-rejecting Polyakov
shift updates are alternated shows pronounced critical slowing down. The shown autocorrelation times
are in units of sweeps whose cost is proportional to the lattice volume.

A milder critical slowing down is observed if cluster updates are included. A cluster
update was carried out whenever after a Polyakov-shift $\mathbf u$ and $\mathbf v$ coincided.
Again sweeps were constructed such that their cost scales like $L^D$ (this requires knowledge about
the average cluster size).

\subsection{In the Confined Phase}
In the confined phase, $\beta<\beta_c$, an interesting quantity is the 
closed-string massgap and its dependence on the temporal extent $L_0$.
It is extracted from the large $x_1$ behavior of 
zero transverse momentum correlators
$C(x_1) = \sum\limits_{x_2} G(\mathbf{x})$. These are measured on
lattices of size $L_0\times L\times L$ for a range of $L_0$ values.

\begin{figure}[h]
   \begin{center}
   \includegraphics[width=0.7\linewidth]{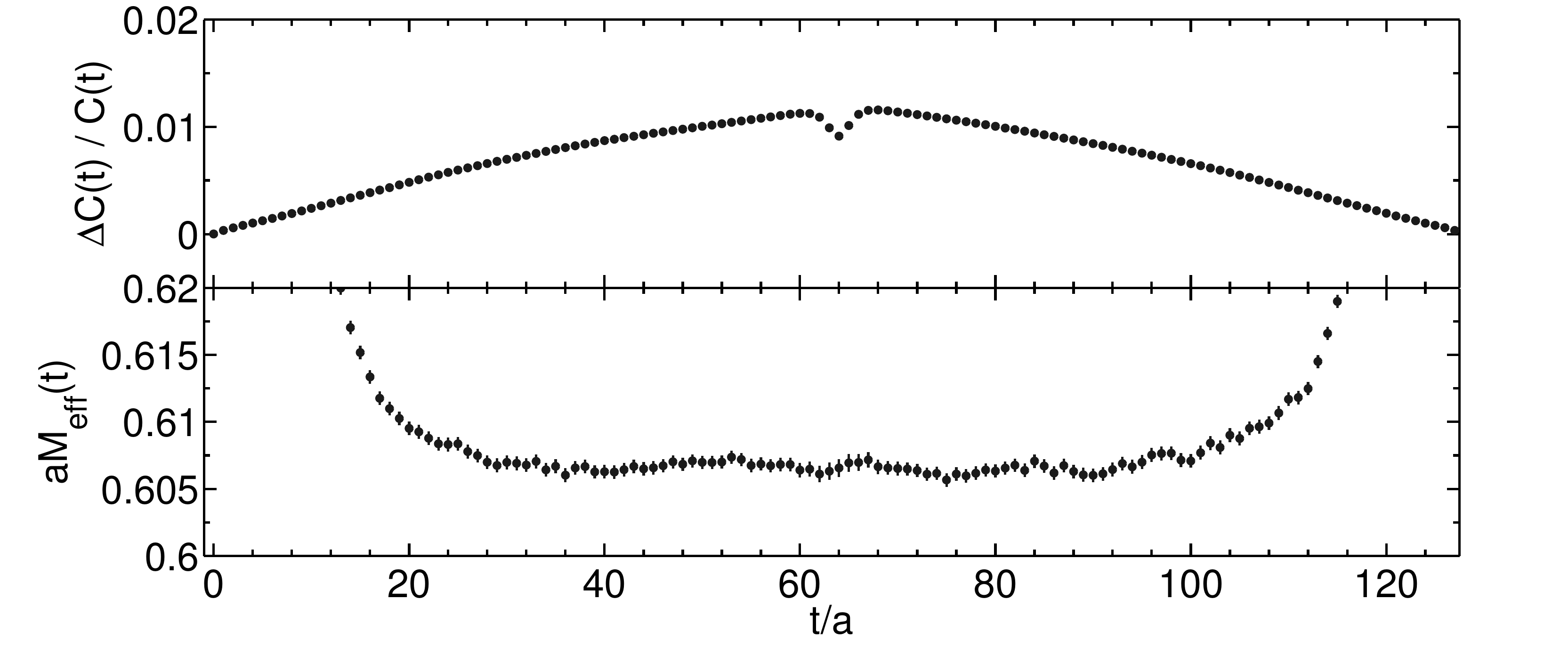}
   \end{center}
   \caption{Extraction of the closed string mass gap. The relative errors of the zero momentum Polyakov line correlator (top)
   and the effective mass (bottom) are shown as a function of separation in the $\hat 1$
   direction.}\label{FIGmeff}
\end{figure}

To achieve a good signal/noise ratio at large separations, we exploit our 
freedom to choose $\rho$ and set it to
\begin{equation}
   \rho(\mathbf{x}) = \sum\limits_{\mathbf k \in \mathbb{Z}} K_0\left(\hat M 
            \sqrt{(x_1-k_1L)^2+ (x_2-k_2L)^2}\right)\, .
\end{equation}
The r.h.s captures the main features of the expected correlation function and leaves
us with a single algorithmic parameter $\hat M$. $K_0$ is the modified Bessel function.
Fig.~\ref{FIGmeff} shows a typical result. A fit to the effective mass
$a M_\text{eff}(t) = {\rm acosh}\left[\frac{C(t-a)+C(t+a)}{2C(t)} \right]$ in the plateau 
region in this case yields $aM=0.60661\pm 0.00024$, which is a very accurate result
for this type of quantity.
A detailed study of the closed string massgap and a comparison with predictions from  effective 
string theories based on our algorithm is reported on
in~\cite{Korzec:2013tia}.

\section{Conclusions}
Simulating the all order strong coupling expansion of an enlarged ensemble has almost completely removed 
critical slowing down in various spin systems. This simple recipe so far does not appear to carry over to gauge theories. 
Other aspects of worm algorithms, in particular the ability to sample two point functions
with (exponentially) increased precision have been successfully transferred to the present case.

\bibliography{mainz}{}
\bibliographystyle{JHEP-2}

\end{document}